# The Journal Impact Factor:
# A brief history, critique, and discussion of adverse effects

Vincent Larivière[1,2] & Cassidy R. Sugimoto[3]

[1] École de bibliothéconomie et des sciences de l'information, Université de Montréal, Canada.
[2] Observatoire des sciences et des technologies (OST), Centre interuniversitaire de recherche sur la science et la technologie (CIRST), Université du Québec à Montréal, Canada.
[3] School of Informatics and Computing, Indiana University Bloomington, USA.

**Table of Contents**

**Abstract**
The Journal Impact Factor (JIF) is, by far, the most discussed bibliometric indicator. Since its introduction over 40 years ago, it has had enormous effects on the scientific ecosystem: transforming the publishing industry, shaping hiring practices and the allocation of resources, and, as a result, reorienting the research activities and dissemination practices of scholars. Given both the ubiquity and impact of the indicator, the JIF has been widely dissected and debated by scholars of every disciplinary orientation. Drawing on the existing literature as well as on original research, this chapter provides a brief history of the indicator and highlights well-known limitations—such as the asymmetry between the numerator and the denominator, differences across disciplines, the insufficient citation window, and the skewness of the underlying citation distributions. The inflation of the JIF and the weakening predictive power is discussed, as well as the adverse effects on the behaviors of individual actors and the research enterprise. Alternative journal-based indicators are described and the chapter concludes with a call for responsible application and a commentary on future developments in journal indicators.

**1. Introduction**
In the 1975 version of the Science Citation Index (SCI), Eugene Garfield and the Institute for Scientific Information (ISI) added a new component to their information products: the Journal Citation Reports (JCR). While Garfield and Sher proposed the concept of an *impact factor* as early as 1963—and tested it at a larger scale in 1972 (Garfield, 1972)—the 1975 JCR was ISI's first comprehensive reporting of their data at the journal level. Based on more than 4.2 million references made in 1974 by 400,000 papers published in about 2,400 journals, this new information source provided a detailed list of journal-to-journal citation linkages, as well as the first iteration of what would become the most discussed and derided bibliometric indicator: the Journal Impact Factor (JIF). (For a detailed history of the Journal Impact Factor see Archambault and Larivière (2009).)

Garfield did not leave the community without a roadmap. In two short papers introducing the first edition of the JCR—entitled *I. Journals, References and Citations*, and *II. Why the Journal Citation Reports*—Garfield provides words of both caution and optimism. Replying to some of the criticism leveled at the Science Citation Index from the scientific community, he provided a justification for interpreting citations as indicators of the *usage* of scholarly literature: "The more frequently a journal's articles are cited, the more the world's scientific community implies that it finds the journal to be a carrier of useful information" (Garfield, 1976b, p. 1). Understanding usage, wrote Garfield, would provide critical information on the economics of scholarly publishing and help librarians "counteract the inertia that too often prevails with regard to journal selection" (p. 1). Data contained in the JCR would, Garfield argued, provide objective indicators for the use of journals so that librarians could make timely and informed decisions on collection management. The report would provide at scale what had required painstakingly manual analyses in previous decades (e.g., Gross & Gross, 1927). For researchers, Garfield imagined that the JCR would help them to identify potential venues for publication. Garfield did not advocate for using the JCR to identify elite journals. Rather, he suggested that researchers use the journal-to-journal matrix to identify multidisciplinary venues at "the borders of their own fields". Garfield (1976c, p. 4-5) writes:

> "… the JCR© can be very helpful in deciding where to publish to reach the audience you want to reach. If, for example, you have a paper that deals with some interesting mathematical aspects of biological problems but is nevertheless definitely a biological paper, the JCR© show you which biological journals have the best 'connections' with math, and which are most likely to welcome the paper."

Furthermore, Garfield saw in these new reports the potential to uncover many important dimensions about the nature of science itself. In the conclusion of the introduction to the JCR, Garfield states (1976c, p. 5):

> "The use of the JCR can be of far-ranging significance in a field about which I can say least here--science--its planning, its evaluation, its sociology, its history. Citation analysis can be used to identify and map research fronts; to define disciplines and emerging specialties through journal relationships; to determine the interdisciplinary or multidisciplinary character and impact of research programs and projects. I say least about this, to me the most exciting aspect of its potential, because the JCR in its present form is, for such advanced applications, only a sketch of that potential, providing little more than suggestions for further and deeper examination of the massive data bank from which its sections have been extracted."

Garfield concludes with a statement of his hopes: that the JCR will "provide material for innovative research", prompting "imaginative analyses", and stimulate "with every answer it gives more questions that need answers" (Garfield, 1976c, p. 5). Along these lines, Garfield writes in the preface of the first JCR:

> "In the introduction I have tried to explain clearly what the JCR is, how it was compiled, how it can be used for some simple purposes for which, I think, it is certainly needed. I have tried also to suggest its usefulness in what I'll call more advanced research. If I have failed in the latter, it is because I have deliberately, and with some difficulty, restrained my own enthusiasm about the value of what some may find at first sight to be merely another handbook of data. Let me say only that the sociology of science is a relatively new field. I believe that JCR will prove uniquely useful in exploring it" (1976a, p. I).

The JCR did indeed provoke a reaction within the research community. Spurred by Derek de Solla Price's call for a *science of science* (Price, 1963), scholars turned to the ISI for data. The JCR and associated products became the backbone for the burgeoning field of scientometrics which sought to address, quantitatively, the questions of science: "its planning, its evaluation, its sociology, its history". In addition to fueling science studies, the JCR found new application alongside the growing emphasis on research evaluation as scholars, institutions, policy-makers, and publishers sought to find ways to measure the success of the research enterprise. This, in turn, had sizeable effects on the science system and scholarly publishing, orienting scholars' research topics and dissemination practices, as well as universities' hiring practices (Monastersky, 2005; Müller & De Rijcke, 2017).

The primary indicator of the JCR—the Journal Impact Factor (JIF)—has received global attention. As of August 2017, the Core Collection of the Web of Science contained more than 5,800 articles that mention the JIF. These papers are not solely in the domain of information or computing science; rather, the majority of papers dealing with JIF are published in scientific and medical journals, demonstrating the pervasive interest in this indicator across scientific fields. The goal of the present chapter is not to summarize this literature *per se*, but rather to focus on the central limitations that have been raised in the literature and among members of the scientific community.

Drawing on the existing literature as well as on original data, this chapter provides an overview of the JIF and of its uses, as well as a detailed, empirically-based, discussion of common critiques. These include technical critiques—such as the asymmetry between the numerator and the denominator, the inclusion of journal self-citations, the length of the citation window, and the skewness of citation distributions—and interpretative critiques—such as the field- and time-dependency of the indicator. Adverse effects of the JIF are discussed and the chapter concludes with an outlook on the future of journal-based measures of scientific impact.

**2. Calculation and reproduction**
The calculation of the JIF is relatively straightforward: the ratio between the number of citations received in a given year by documents published a journal during the two previous years, divided by the number of items published in that journal over the two previous years. More specifically, the JIF of a given journal for the year 2016 will be obtained by the following calculation:

Number of citations received in 2016 by items published in the journal in 2014-2015
divided by
Number of citable items published in the journal in 2014-2015

Citable items are restricted, by document type, to articles and reviews in the denominator, but not in the numerator (McVeigh & Mann, 2009); an issue we will discuss more in-depth later in the chapter. Therefore, the JIF is generally interpreted as the mean number of citations received by papers published in a given journal in the short term, despite not being exactly calculated as such.

Given its calculation, which uses one year of citation and two years of publication, it combines citations to papers that have had nearly three years of potential citations (i.e., papers published in early 2014) with citations to papers which have had slightly more than a year to receive citations (i.e., papers published at the end of 2015). The JIF is presented with three decimals to avoid ties. However, this has been argued as "false precision" (Hicks, et al., 2015) with critics advocating for the use of only one decimal point.

Each journal indexed by Clarivate Analytics in the Science Citation Index Expanded (SCIE) and the Social Science Citation Index (SSCI) receives an annual JIF. Given the long half-life of citations (and references) of journals indexed in the Arts and Humanities Citation Index (AHCI), these journals are not provided with a JIF (although some social history journals indexed in the SSCI are included). There has been a steady increase in the number of journals for which JIFs are compiled, in parallel with the increase in indexation. In 1997, 6,388 journals had JIFs. This number nearly doubled 20 years later: in 2016, 11,430 received a JIF.

Despite the apparent simplicity of the calculation, JIFs are largely considered non-reproducible (Anseel et al., 2004; Rossner, Van Epps, Hill, 2007). However, in order to better understand the calculation of the JIF, we have attempted to recompile, using our licensed version of the Web of Science Core Collection (which includes the Science Citation Index Expanded, Social Science Citation Index, and Arts and Humanities Citation Index), the 2016 JIFs for four journals from the field of biochemistry and molecular biology (*Cell*, *Nature Chemical Biology*, *PLOS Biology,* and *FASEB J*).

We begin with a careful cleaning of journal names to identify citations that are not automatically matched in WOS—that is, citations that bear the name of the journal, but contain a mistake in the author name, volume, or number of pages. The inclusion of these unmatched citations provides the opportunity to essentially reverse-engineer the JIFs presented in the JCR. This reduces the opacity of the JCR, which many consider to be the results of calculations performed on a "separate database" (Rossner, Van Epps, Hill, 2007).

Our empirical analysis (Table 1) shows that the inclusion of unmatched citations and the variants under which journal names appear (WOS-derived JIF) provides results that are very similar to the official JCR JIF. This suggests that there is no separate database and one can closely approximate the JIF using only the three standard citation indexes contained the Core Collection. Furthermore, our results suggest that papers indexed in Clarivate's other indexes—e.g., the Conference Proceedings Citation Index and Book Citation Index—are not included. The inclusion of these databases would lead to an increase of the JIF for most journals, particularly those in disciplines that publish a lower proportion of their work in journals. Most importantly, our analysis demonstrates that with access to the data and careful cleaning, the JIF can be reproduced.

Table 1. Citations received, number of citable items, WOS-derived JIF, JCR JIF and proportion of papers obtaining the JIF value, for four journals from the field of biochemistry and molecular biology, 2014-2015 papers and 2016 citations

| Journal | Citations | | All Citations | N. Citable Items | WOS-derived JIF | JCR JIF |
|---|---|---|---|---|---|---|
| | Matched items | Unmatched items | | | | |
| Cell | 24,554 | 2,016 | 26,570 | 869 | 30.575 | 30.410 |
| Nat. Chem. Biol. | 3,858 | 356 | 4,214 | 268 | 15.724 | 15.066 |
| PLOS Biol. | 3,331 | 290 | 3,621 | 384 | 9.430 | 9.797 |
| FASEB J. | 4,088 | 802 | 4,890 | 881 | 5.551 | 5.498 |

## 3. Critiques

The JIF has been called a "pox upon the land" (Monastersky, 2005), "a cancer that can no longer be ignored" (Curry, 2012), and the "number that's devouring science" (Monastersky, 2005). Many scholars note the technical imperfections of the indicator—skewness, false precision, absence of confidence intervals, and the asymmetry in the calculation. Considerable focus has also been paid to the misapplication of the indicator—most specifically the use of the indicator at the level of an individual paper or author (e.g., Campbell, 2008). We will not review this vast literature here, much of which appears as anecdotes in editorial and comment pieces. Instead, we provide original data to examine the most discussed technical and interpretive critiques of the JIF. Furthermore, we provide new information on a previously understudied dimension of the JIF—that is, the inflation of JIFs over time.

### 3.1 The numerator / denominator asymmetry

Scholarly journals publish several document types. In addition to research articles, which represent the bulk of the scientific literature, scholarly journals also publish review articles, which synthesize previous findings. These two document types, which are generally peer-reviewed, account for the majority of citations received by journals and constitute what Clarivate labels as "citable items". Over the 1900-2016 period, 69.7% of documents in the Web of Science were considered as citable items. This proportion is even more striking for recent years, with 76.0% of documents published in 2016 labeled as citable items. Other documents published by scholarly journals, such as editorials, letters to the editor, news items, and obituaries (often labelled as "front material"), receive fewer citations, and are thus considered "non-citable items". There is, however, an asymmetry in how these document types are incorporated into the calculation of the Journal Impact Factor (JIF): while citations received by all document types—citable and non-citable—are counted in the numerator, only citable items are counted in the denominator. This counting mechanism is not an intentional asymmetry, but rather an artifact of method for obtaining citation counts. As mentioned above, to account for mistakes in cited references and to try to be as comprehensive as possible, Clarivate focuses retrieval on all citations with the journal name or common variant (Hubbard & McVeigh, 2011) rather than using a paper-based approach to calculating citations. This has the effect of inflating the JIF: citations are counted for documents which are not considered in the denominator. The variations in document types (i.e., reduction of the number of citable items in the denominator) has also been argued as the main reason for JIF increases (Kiesslich, Weineck, Koelblinger, 2016).

To better understand the effects of document types on the calculation of the JIF, we compiled, for the sample of four journals from the field of biochemistry and molecular biology, as well as for *Science* and *Nature*—both of which publish a high percentage of front material— citations received by citable items, non-citable items, as well as unmatched citations (Table 2). Following Moed and van Leeuwen (1995a, 1995b), our results show that non-citable items and unmatched citations account for a sizeable proportion of total citations received, from 9.8% in the case of *Cell* to 20.6% in the case of *FASEB Journal*. For the four journals from biochemistry and molecular biology, unmatched citations account for a larger proportion of citations than non-citable items. Given that these unmatched citations are likely to be made to citable items, this suggests that, at least in the case of disciplinary journals which do not typically have a large proportion of front material, the asymmetry between the numerator and the denominator does not inflate JIFs in a sizeable manner. The effect of non-citable items is much greater for interdisciplinary journals such as *Science* and *Nature*. As shown in Table 2, for both *Nature* and *Science*, more than 5,000 citations are received in 2016 by non-citable items published in the journal in 2014-2015. This accounts for 7.2% and 9.0% of citations, respectively, which is greater than the percentages obtained by the sample of disciplinary journals [2.3%-6.5%]. Results also show that the difference in the "symmetric" JIF—with only citable items in the numerator and denominator—and JCR JIF is greater for *Nature* and *Science* than *Cell* or *Nat. Chem. Biol.*, mostly because of citations to non-source items. However, at scale—i.e., all journals having a JIF in 2016—the relationship between the JIF and the symmetric Impact Factor is quite strong, with an $R^2$ of 0.96 (Figure 1).

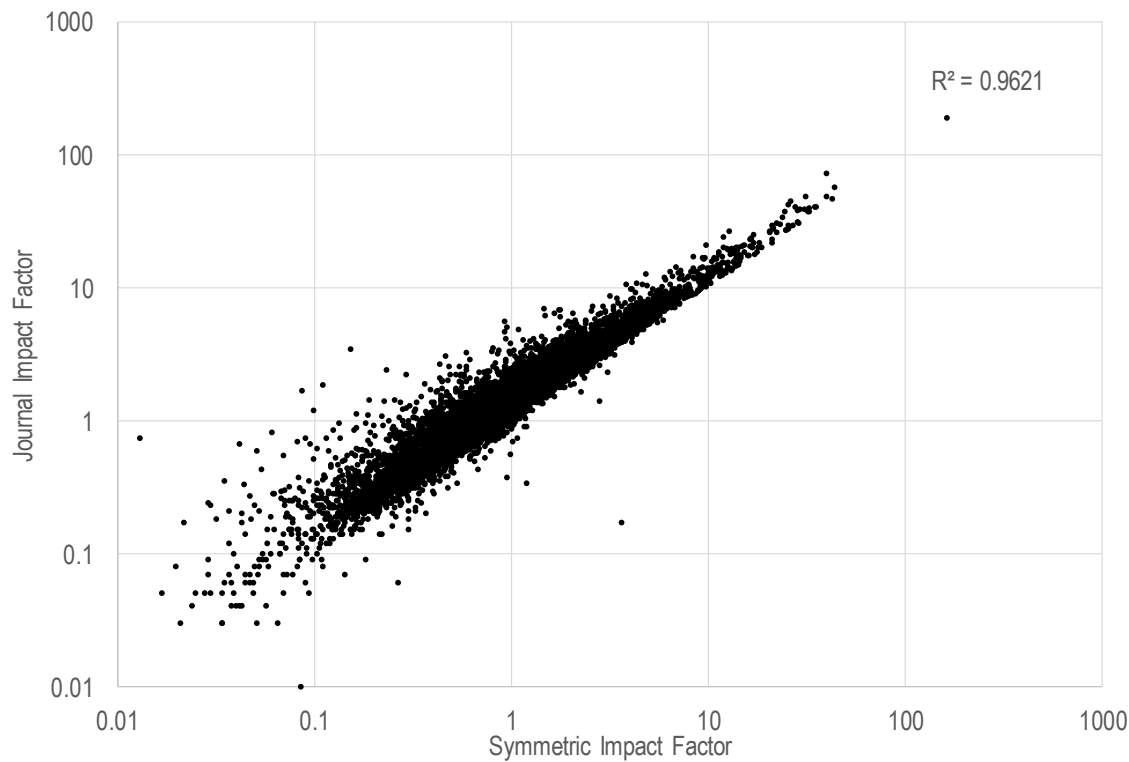

Figure 1. Correlation between the JIF and the symmetric Impact Factor, 2016

These results demonstrate that the asymmetry has different effects based on 1) the proportion of front material, and 2) the completeness of citations received by the journal. Moreover, they show that most of the additional citations—i.e., citations not directly linked to citable items—are unmatched citations rather than direct citations to non-citable items. Given most of these unmatched citations are likely to be directed at source items, a more accurate calculation of the JIF could exclude citations to non-source items, but retain unmatched citations. Of course, the ideal solution would be to perform additional data cleaning to reduce the proportion of unmatched citations and have perfect symmetry between the numerator and denominator.

Table 2. Number and proportion of citations received by articles, reviews, non-citable items, and unmatched citations, for four journals from the field of biochemistry and molecular biology, as well as *Nature* and *Science*, 2014-2015 papers and 2016 citations

| Journal | Article | | Review | | Non-Citable Items | | Unmatched Citations | | N. Citable Items | Symmetric Impact Factor | JCR Impact Factor | % Increase |
|---|---|---|---|---|---|---|---|---|---|---|---|---|
| | N | % | N | % | N | % | N | % | | | | |
| Cell | 20,885 | 78.6% | 3,068 | 11.5% | 601 | 2.3% | 2,016 | 7.6% | 869 | 27.564 | 30.410 | *10.3%* |
| Nat. Chem. Biol. | 3,263 | 77.4% | 378 | 9.0% | 217 | 5.1% | 356 | 8.4% | 268 | 13.586 | 15.066 | *10.9%* |
| PLOS Biol. | 3,088 | 85.3% | 6 | 0.2% | 237 | 6.5% | 290 | 8.0% | 384 | 8.057 | 9.797 | *21.6%* |
| FASEB J. | 3,650 | 74.6% | 235 | 4.8% | 203 | 4.2% | 802 | 16.4% | 881 | 4.410 | 5.498 | *24.7%* |
| Nature | 55,380 | 78.6% | 3,925 | 5.6% | 5,067 | 7.2% | 6,047 | 8.6% | 1,784 | 33.243 | 40.140 | *20.7%* |
| Science | 45,708 | 73.0% | 4,886 | 7.8% | 5,657 | 9.0% | 6,340 | 10.1% | 1,721 | 29.398 | 37.210 | *26.6%* |

## 3.2 Journal self-citations

The inclusion of journal self-citations in the calculation of the Journal Impact Factor (JIF) has been a cause for concern, as it opens the door for editorial manipulations of citations (Arnold & Fowler, 2011; Reedijk & Moed, 2008; Martin, 2013). Journal self-citations are those citations received by the journal that were made by other papers within that same journal. This should not be conflated with self-references, which is the proportion of references made in the articles to that journal. This is a subtle, but important difference: the proportion of self-citations is an indication of the relative impact of the work on the broader community, whereas the proportion of self-references provides an indication of the foundation of work upon which that journal is built. From a technical standpoint, the main concern in the construction of the JIF is the degree to which self-citations can be used to inflate the indicator. Given that self-citations are directly under the control of the authors (and, indirectly, the editors), this has been seen as a potential flaw that can be exploited by malicious authors and editors.

There are many myths and misunderstandings in this area. For example, it has been argued that authors in high impact journals are more likely to self-cite than those in low-impact journals because "the former authors in general are more experienced and more successful" (Anseel et al., 2004, p. 50). However, this is a conflation of self-citations and self-references. Authors with longer publication histories are, indeed, more likely to have material to self-reference. However, successful authors are likely to have lower self-citation rates, as they are likely to generate citations from a broader audience. Furthermore, this conflates the practices of an individual author (who publishes in many journals) to the self-citation of a journal, which is much more dependent upon the specialization of the journal, among other factors (Rousseau, 1999). There is also a distinction to made between the number and proportion of self-citations. As ISI observed in internal analyses, "a high number of self-citations does not always result in a high rate of self-citation" (McVeigh, 2002, par. 15). For example, a study of psychology journals found that articles in high-impact journals tend to receive a higher *number* of self-citations than articles in lower-impact journals; however, the *ratio* of self-citations to total citations tends to be lower for high-impact journals (Anseel et al., 2004).

Producers of the JIF thus face a Cornelian dilemma when it comes to self-citations: while including them can lead to manipulation, excluding them penalizes niche journals and certain specialties. In response to these concerns, ISI undertook an analysis of the prevalence and effect of journal self-citations (McVeigh, 2002). In an analysis of 5,876 journals in the 2002 Science Edition of the JCR, ISI found that the mean self-citation rate was around 12%. Our analysis of 2016 citation data for papers published in 2014-2015 reinforces this: we find that the percentage of self-citations across all disciplines remains around 12% (Figure 2). However, the percentage varies widely by discipline, with Arts and Humanities having far higher degrees of self-citation than Clinical and Biomedical research. This suggests that, on average, the majority of citations do not come in the form of self-citations and makes abuses easier to identify.

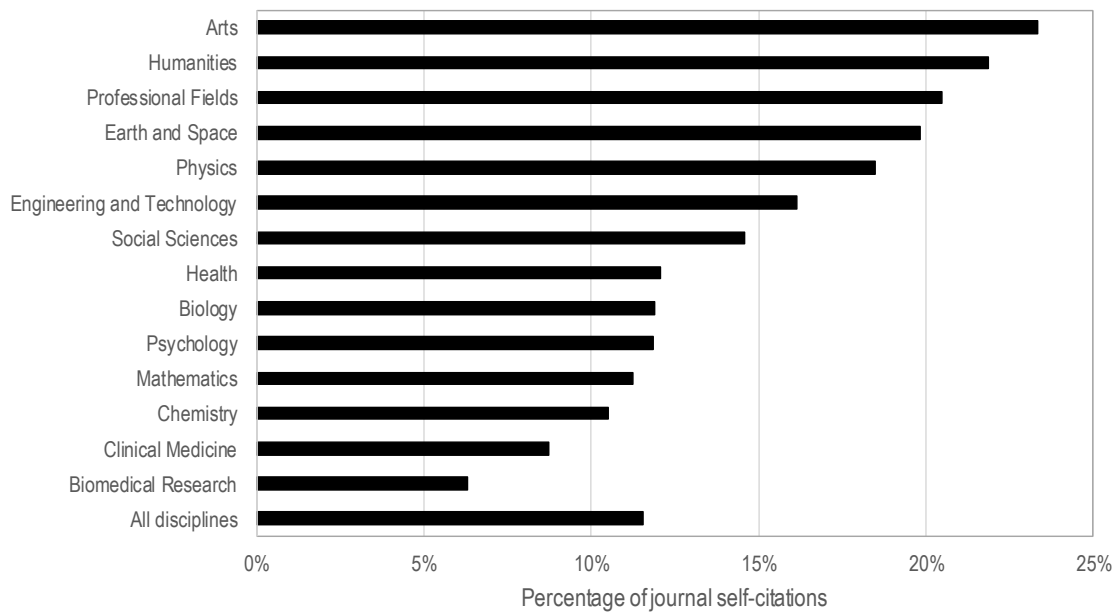

Figure 2. Percentage of journal self-citations, by discipline, for citations received in 2016 by papers published in 2014-2015

The ISI analysis also examined the correlation between self-citation rates and JIFs. While studies focusing on particular domains have found varying results (e.g., Nisonger, 2000; Fassoulaki et al, 2000; Anseel et al., 2004; Opthof, 2013), the large-scale analysis by ISI found a weak negative correlation between JIF and rates of journal self-citation (McVeigh, 2002). The analysis noted that self-citation had little effect on the relative ranking of high impact journals, given that journals in the top quartile of JIFs tended to have self-citation rates of 10% or less. Lower impact journals, however, were more dependent upon self-citations (McVeigh, 2002). We found similar results for all 2016 journals. As shown in Figure 3, there is a relatively strong correlation between a journal's total number of external citations (i.e., non self-citations) and its number of self-citations (left panel), which suggests that external- and self-citations are related, but also that there are other factors influencing the relationship, such as the level of specialism of the journal. For instance, 2014-2015 papers from the *Journal of High Energy Physics* received 18,651 citations in 2016, of which 9,285 (50%) came from the same journal. Other more generalist journals in that domain—such as *Physical Review B* and *Monthly Notices of the Royal Astronomical Society*—exhibit a similar pattern.

The irony of the concern between self-citation and JIFs, however, is that the relationship is inverted: there is actually a negative relationship between the percentage of self-citations for a journal and the JIF (Figure 3, left panel). That is, those journals with the highest JIFs tend to have the lowest percentage of self-citations. There is, simply speaking, a limit on the advantages of self-citations. There are many more articles outside of the journal than within and relying on citations within can only generate a finite number of citations. A variant JIF omitting self-citations is now available in the JCR. However, the two-year JIF including self-citations continues to be the dominant form.

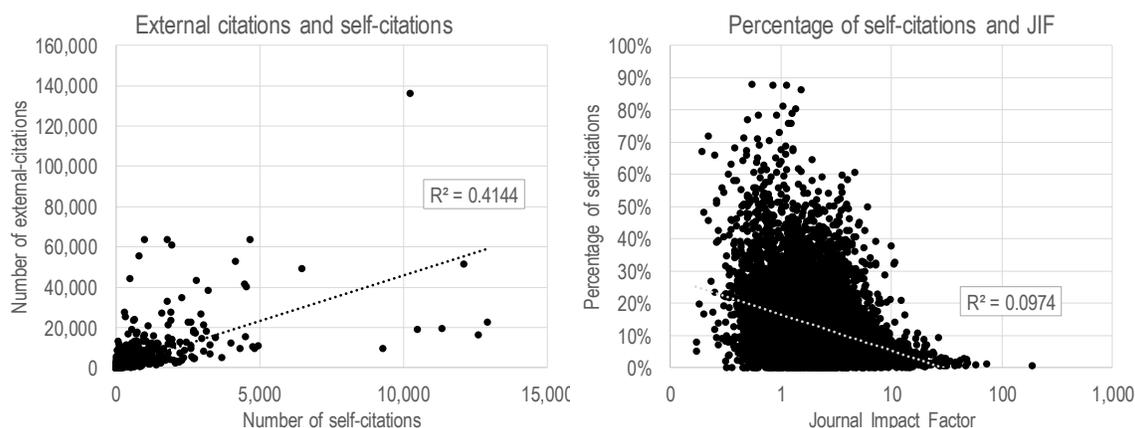

Figure 3. Correlation at the journal level between A) the number of journal "external" citations and number of journal self-citations and B) the percentage of self-citations and the Impact Factor, for year 2016. Only journals with at least 50 citations in 2016 to material published in 2014-2015 are shown.

**3.3 Length of citation window**

The Journal Impact Factor (JIF) includes citations received in a single year by papers published in the journal over the two preceding years. As such, it is generally considered to cover citations received by papers over a two-year window. This focus on the short-term impact of scholarly documents is problematic as it favors disciplines that accumulate citations faster. For example, comparing mean citation rates of papers published in the *Lancet* and in the *American Sociological Review* (ASR)—two journals with very different JIFs (47.83 vs 4.4 in 2016)—Glänzel and Moed (2002) have shown that while papers published in the *Lancet* had a higher mean citation rates for two- and three-year citation windows, those published in *ASR* were more highly cited when a longer citation window was used.

This trend can be observed at the macro-level: Figure 4 presents the annual number of citations (left panel), cumulative number of citations (middle panel), and the cumulative proportion of citations (right panel), for all papers published in 1985 across four disciplines (Biomedical Research, Psychology, Physics, and Social Sciences). These data show that citations to Biomedical Research and Physics peaks two years following publications, while citations are relatively more stable following publication year in Psychology and the Social Sciences. It is particularly revealing that Psychology papers receive, on average, more citations (cumulatively) than Physics papers. While Physics papers generate more citations than Psychology papers within the first five years, the reverse is true for the following 25 years.

Despite these disciplinary differences in the speed at which citations accumulate, the two-year window appears to be ill-suited across all disciplines, as it covers only a small fraction of citations received over time. For example, using a 30-year citation window, we find that the first two years captures only 16% of citations for Physics papers, 15% for Biomedical Research, 8% for Social Science papers, and 7% in Psychology. Figure 4 also shows that papers in Biomedical Research accumulate citations faster than in the other three domains. For instance, they accumulate 50% of their citations in the first eight years following publication, while it takes nine years for Physics papers, 13 years for Psychology papers, and 14 years for Social Science papers to reach the same threshold. In order to take such differences into account, the JCR has provided, since 2007, a 5-year JIF. Despite this improved citation window, which provides a more complete measurement of the impact of papers and journals, the two-year JIF remains the gold standard.

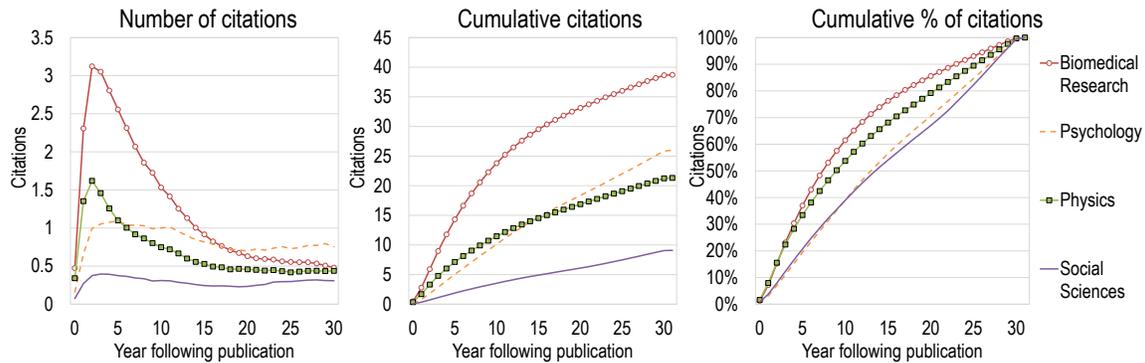

Figure 4. Number of citations (left panel), cumulative number of citations (middle panel), and cumulative proportion of citations (right panel), by year following publication for papers published in 1985 in biomedical research, psychology, physics and social sciences (NSF classification).

### 3.4 Skewness of citation distributions

Nearly a century of research has demonstrated that science is highly skewed (Lotka, 1926) and that productivity and citedness are not equally distributed among scholars, articles, institutions, or nations. It is perhaps of little surprise, therefore, that the citedness of articles within a journal is also highly skewed. This was the main premise of an article published in 1992 by Per O. Seglen, who produced a robust empirical analysis demonstrating that a minority of papers in a journal accounted for the vast majority of citations. Given this skewness in the citation distribution, Seglen (1992) argued that the JIF was unsuitable for research evaluation.

To illustrate this skewness, we provide—for the four biochemistry and molecular biology journals mentioned above—the distribution of citations received in 2016 by papers published in 2014-2015, both as absolute values (Figure 5) and as percentages of papers (Figure 6). It shows that, for all journals, most of the papers have a low number of citations and only a few obtain a high number of citations. Of course, the distribution for *Cell*—with a JIF of 30.410—is more right-skewed than *FASEB J.*—which has a JIF of 5.498—but despite this, their citation distributions still have sizeable overlap, as shown in Figure 6. Also striking is the similarity of the skewness: for all of these four journals, a nearly identical percentage of papers—28.2%-28.7%—obtain a citation rate that is equal or greater to the JIF for that journal.

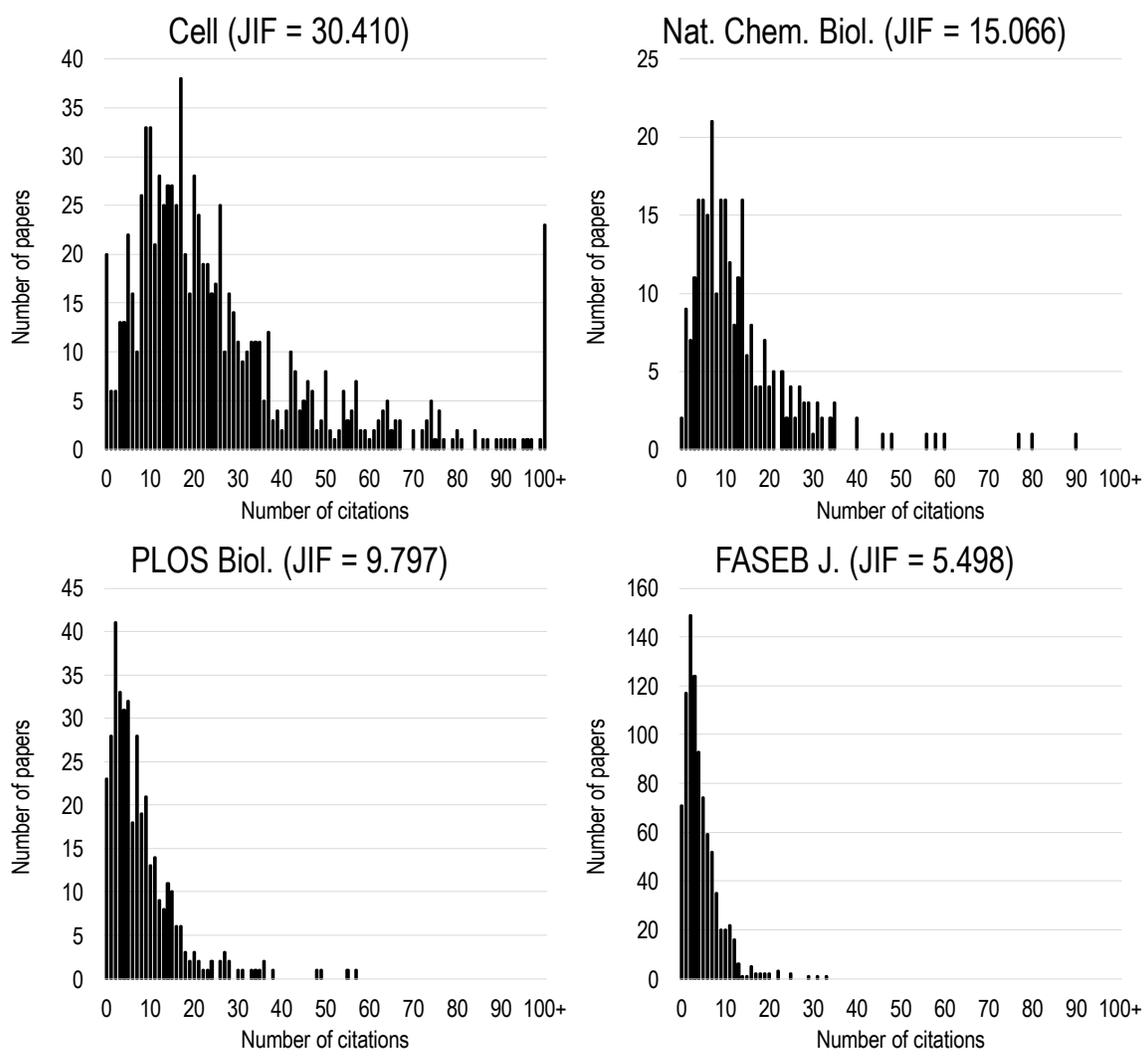

Figure 5. Distribution of citations received by articles and reviews, for four journals from the field of biochemistry and molecular biology, 2014-2015 papers and 2016 citations

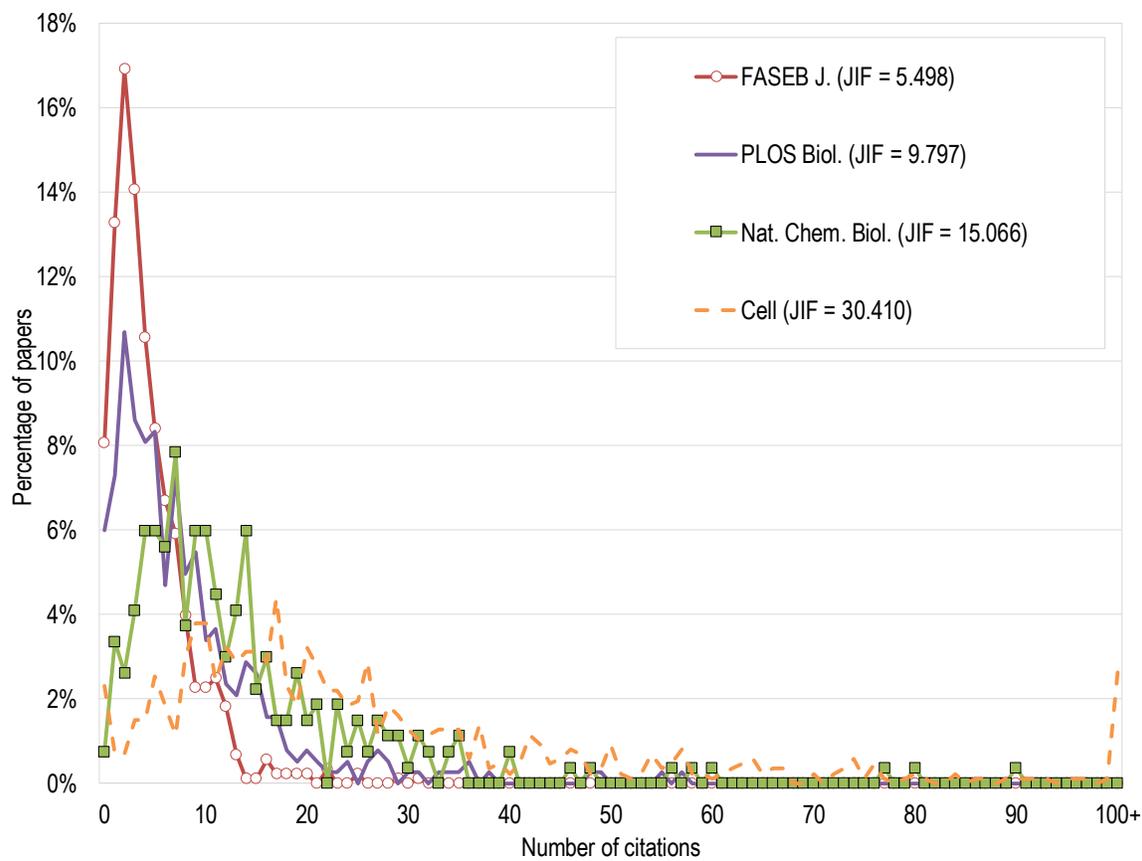

Figure 6. Relative distribution of citations received by articles and reviews, for four journals from the field of biochemistry and molecular biology, 2014-2015 papers and 2016 citations

Extending the analysis across all journals indexed in the 2016 JCR confirms this pattern (Figure 7). There is a fairly normal distribution when plotting journals by the percentage of their papers that obtain the corresponding JIF value or above. As shown, the vast majority are around 30%. Nearly 73% of the journals fall between 20-40%. Only in 1.3% of journals (n=141) do at least 50% of the articles reach the JIF value. This fundamental flaw in the calculation—to compile an average on a non-parametric distribution—has been heavily discussed in the literature (Larivière et al., 2016) as both a statistical aberration and also for the common misinterpretation: to use the JIF as an indicator at the article or individual level. Our analysis demonstrates the fairly weak predictive power of the JIF—that is, one cannot extrapolate from the impact factor of the journal to the potential citedness of the article as only one-third of the articles are likely to obtain that value.

It would be irresponsible here not to mention the Lucas Critique (Lucas, 1976), which argues against predicting the effects of policy changes based on aggregated historical data. The Lucas Critique was developed for economic data, but has wide applicability for the social sciences. In bibliometrics, one should be wary of making predictions about future citations, based on the past performance of scholarly objects. Referencing and citing patterns vary over time as do the socio-political factors of scholarship. Furthermore, the construction of citation indicators changes behavior (as we discuss later in this chapter). Therefore, we caution against making predictions with citation data.

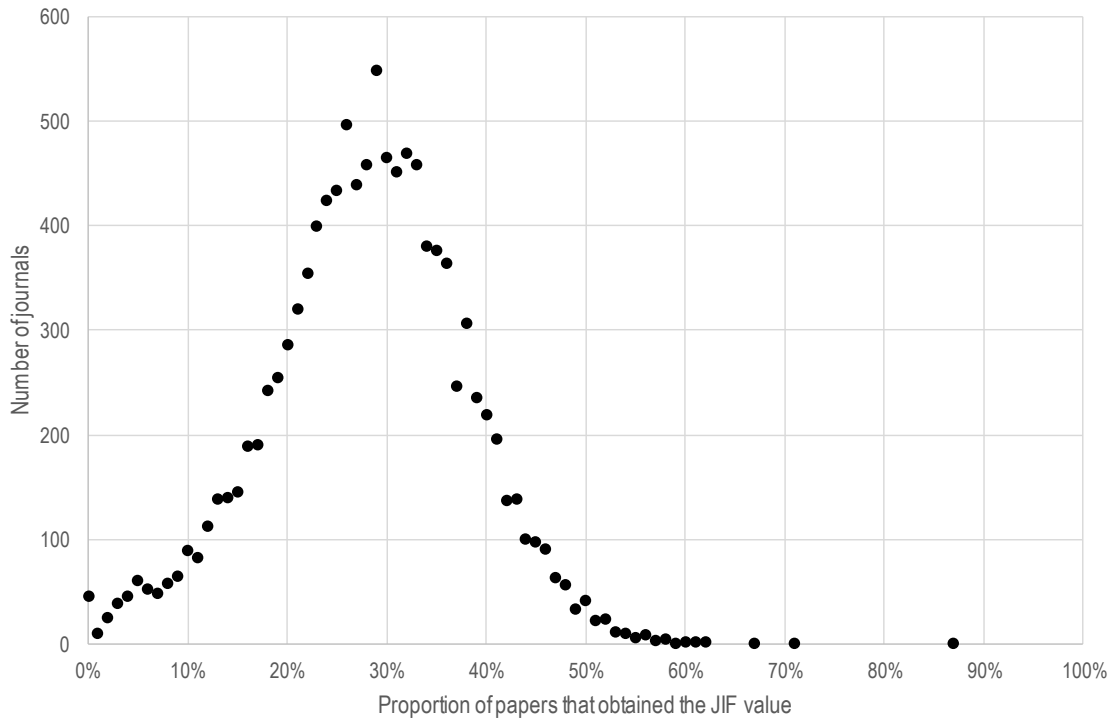

Figure 7. Distribution of the number of journals, by proportion of papers that obtained the JIF value, 2014-2015 papers and 2016 citations

This is not to say, of course, that there is no relationship between JIF and future citedness. For example, using identical papers published in journals with different JIFs, Larivière and Gingras (2010) found that the mean number of citations of the paper published in the journal with the highest JIF obtained twice as many citations as its twin published in the journal with the lowest JIF. However, the relationship between the JIF and the citedness of the articles has weakened over time: as shown by Lozano, Larivière and Gingras (2011) using Web of Science data—and confirmed by Acharya (2014) using Google Scholar—the correlation between the JIF and article-level citations has been decreasing since the mid-1990s. One potential explanation for this is the changing referencing practices of scholars. Citations are less concentrated over time (Larivière, Gingras, & Archambault, 2009) and scholars are citing increasingly older literature (Larivière, Archambault, & Gingras, 2008) and, as they do, more of the citations fall out of the two-year citation window of the JIF.

There have been many suggestions to account for the skewness, such as compiling a median-based JIF (Sombatsompop, Markpin, & Premkamolnetr, 2004; Rousseau, 2005) or reporting citation distributions (Larivière et al., 2016). However, contrary to other alternatives (such as the 5-year JIF and JIF exlcuding self-citations), no alternatives have been adopted by the JCR to address this limitation.

**3.5 Disciplinary comparison**

Field differences in citations are well established and field-normalized indicators have been the norm for several decades (e.g., Schubert & Braun, 1986; Moed, De Bruin, & van Leeuwen, 1995). However, the JIF is not among these. The simplicity of the calculation fails to normalize for the vast differences in citing practices across disciplines, such as the number of references per document and age of references. As shown in Table 3, disciplines that publish papers with longer cited reference lists—especially in terms of WOS-indexed papers—generally have higher JIFs than those with shorter lists. Furthermore, disciplines that cite more recent material—which fall in the JIF two-year citation window—are more likely to have higher JIFs than those which cite older material.

These differences also highlight the importance of references to other WOS-indexed material (source items), which are those that are taken into account in the compilation of the JIF. For instance, while the mean number of references in Biology and Biomedical Research are almost identical, the mean JIF of journals in Biology is less than half of those in Biomedical Research. This difference is explained by the fact that a large proportion of references made by Biology journals do not count in the calculation of JIF as they are made to non-WOS (and, thus, JCR) material, while the vast majority of references of Biomedical Research journals are to WOS-indexed journals.

Table 3. Mean and maximum JIF of journals, mean number of cited references per paper (all material and only to WOS source items), and mean age of cited literature, by discipline, 2014-2015 papers and 2016 citations

| Discipline | Mean JCR JIF | Maximum JCR JIF | Mean N. Ref. | Mean N. refs. to WOS source items | Mean age of cited literature |
|---|---|---|---|---|---|
| Biology | 1.683 | 22.81 | 48.99 | 34.45 | 14.72 |
| Biomedical Research | 3.526 | 46.6 | 48.94 | 43.19 | 10.26 |
| Chemistry | 2.768 | 47.93 | 46.37 | 41.31 | 10.37 |
| Clinical Medicine | 2.976 | 187.04 | 41.94 | 34.78 | 9.77 |
| Earth and Space | 2.173 | 30.73 | 53.71 | 38.67 | 13.06 |
| Engineering and Technology | 1.989 | 39.74 | 36.35 | 24.77 | 10.44 |
| Health | 1.647 | 17.69 | 39.08 | 24.52 | 9.86 |
| Mathematics | 1.017 | 9.44 | 26.56 | 16.53 | 16.65 |
| Physics | 2.699 | 37.85 | 36.57 | 29.58 | 12.55 |
| Professional Fields | 1.565 | 11.12 | 53.51 | 27.68 | 13.09 |
| Psychology | 2.050 | 19.95 | 54.56 | 38.30 | 13.00 |
| Social Sciences | 1.199 | 6.66 | 49.09 | 21.74 | 15.12 |

The same patterns are observed at the level of NSF specialities (Figure 8). Specialties that cite a higher number of references per paper on average typically have higher JIFs (left panel), as are specialities that cite younger material. Therefore, the indicator cannot be used to compare across disciplines: medical researchers are much more likely to publish in journals with high JIFs than mathematicians or social scientists, and this is strictly due to different disciplines' publication and referencing practices rather than anything that relates to the scholarly impact of the journal.

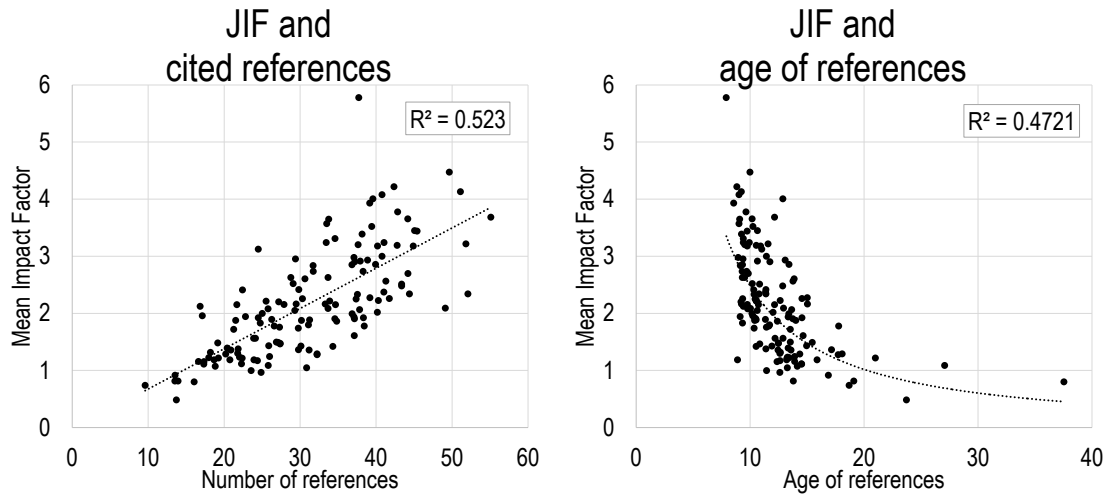

Figure 8. Correlation between the Journal Impact Factor and number of cited references to WOS source items (left panel) and age of references (right panel), by NSF speciality, 2014-2015 papers and 2016 citations

**3.6 Journal Impact Factor inflation**

While the calculation of the JIF has remained stable, values obtained by journals have not. The average JIF value has increased over time, both as a function of the number of papers in existence and the increasing length of their reference lists (Larivière, Archambault, Gingras, 2008). In 1975, the journal with the highest JIF was the *Journal of Experimental Medicine*, with a JIF of 11.874. In the 2016 JCR edition, the highest JIF was 187.040 for *CA: A Cancer Journal for Clinicians*. As shown by Figure 9, a general inflation of the JIF has been observed over the last 20 years. For instance, while only 49 journals (0.8% of total) had JIF above 10 in 1997, this increased to 105 (1.3%) in 2007, and to 201 (1.8%) in 2016. Average JIF values have increased from 1.125 in 1997, to 1.707 in 2007 and then to 2.178 in 2016. Of course, not all journals have observed these increases. One notable example is *PNAS*, which has remained quite stable—the 1975 JIF was 8.989 and, despite some intermittent increases, was only slightly higher at 9.661 in 2016.

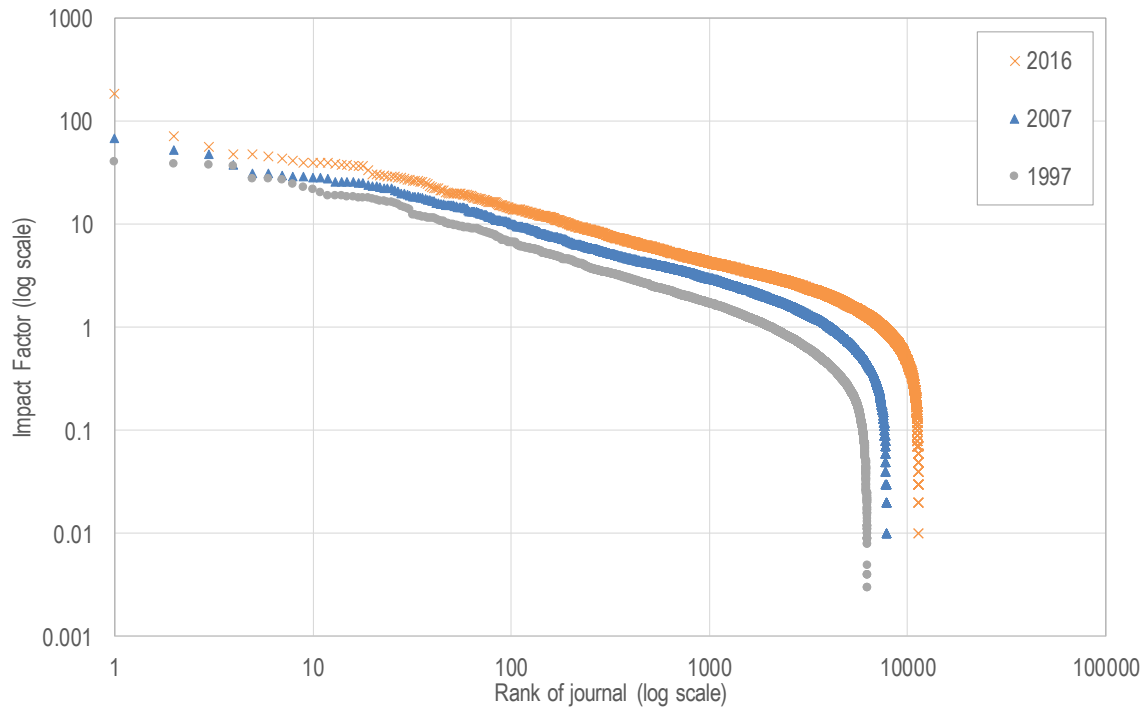

Figure 9. Impact Factor by journal as a function of rank, for years 1997, 2007, and 2016

The inflation of the JIF across time is an important element for interpretation. Many editors wait with baited breath for the release of the next JIF: increases are celebrated as an accomplishment of the editor and the journal (e.g., Bolli, 2017; Govenir, 2016; Simren et al., 2015). Moreover, publishers, such as Elsevier (2007), Springer (2016), and Wiley (2016), among others, publicize their JIF increases with little to no conversation about the expected inflation rates. For example, the Wiley press release boasts that 58% of Wiley journals increased their JIFs between 2014 and 2015. What the press release fails to note is that 56% of all journals in the JCR increased during that same time period. Of course, reporting a relative increase is much less persuasive. As there is no established mechanism for acknowledging inflation in reporting, editors and publishers continue to valorize marginal increased in JIFs which have little relation to the performance of the journal.

**4. Systemic Effects**

There is no doubt that a political economy that has emerged around citation indicators. Nearly two decades ago, Sosteric (1999, p.13) commented on "the neoliberal need for surveillance, the push for administrative measures of scholarly performance and productivity, [and] the growing need for post-publication measures of scholarly impact." He did not characterize scholars as resisters of this panopticon, but rather as adaptive actors in the system. Adaptation for survival and success is well-known across all fields of science: research evaluation is no different. Several scholars have warned against the negative consequences of constructing indicators of social activities (e.g., Campbell, 1979; Goodhart, 1975). As Cronin and Sugimoto summarized (2015, p.751):

> "The use of metrics, whether to monitor, compare or reward scholarly performance, is not a value-neutral activity. Metrics are shaped by, and in turn shape policy decisions; they focus the institutional mind, influence the allocation of resources, promote stratification and competition within science, encourage short-termism and, ultimately, affect the ethos of the academy… As reliance on metrics grows, scholars, more or less consciously, alter the way they go about their business; that is, their behaviors, motivations and values change, incrementally and unwittingly perhaps, as they adapt to the demands and perceived expectations of the prevailing system."

While it would be beyond the scope of the chapter to detail all the systemic effects of scholarly indicators, we focus on the negative and often intentionally malicious effects related to the use and promotion of the JIF. Specifically, we discuss JIF engineering, its relationship with institutional evaluation policies, the application of JIF for evaluating individual researchers and papers, and the creation of imitation indicators.

**4.1 Journal Impact Factor Engineering**

In a context where the JIF determines the fate of a journal—from submission rates to pricing—some editors and publishers have developed subterfuges to increase their JIF which, in turn, decreases the validity of the indicator. Such stratagems aimed at "artificially" increasing impact factors have been called "journal impact factor engineering" (Reedijk & Moed, 2008). One well-documented tactic is to prey on the asymmetry in the calculation and to publish more "front material"—such as editorials, letters to the editor, etc., which are considered by Clarivate as non-citable items (Reedijk & Moed, 2008). Another similar approach is to cite the home journal excessively in editorials and other front matter (Reedijk & Moed, 2008). For example, many journals publish annual "highlights" or other documents with a high number of internal references (Opthof, 2013). Whether malicious or not, these documents unduly inflate—and thereby invalidate—the JIF.

A more subversive approach has been to engage in citation coercion or cartels (Smith, 1997; Monastersky, 2005; Frandsen, 2007; Van Noorden, 2012; Martin, 2013). The expression "citation cartel" is largely attributed to Franck (1999), who used it to refer to the ways in which monopoly power is exercised by publishers and editors on authors in scientific publishing, and noted the complicity of authors who act as "citation-maximizers" in the scholarly communication system. This complicit behavior has been empirically demonstrated: in a study of nearly 7,000 scholars, the majority reported that they would acquiesce to editorial coercion in order to get published (Wilhite & Fong, 2012). The same study also showed that 20% of these scholars said they had been subject to coercive self-citation—that is, requests from editors to add references to irrelevant papers published within the journal (Wilhite & Fong, 2012). An expansion of this study—with new disciplines added—placed this rate at 14.1%. Both the initial and follow-up studied confirmed that coercion was more common among higher impact journals (Wilhite & Fong, 2012; Fong & Wilhite, 2017).

Faced with accusations of extortion (Monastersky, 2005), editors will often argue the innocence of and scientific rationale for these citations (e.g., Cronin, 2012). However, several editors themselves have been caught engaging in JIF boosting, by excessively citing their own journal in editorials (Reedijk & Moed, 2008). There are also egregious examples of coercion. For example, in 2017, the editor of the journal *Land Degradation & Development*—who sat on the board and reviewed for other journals in the field—took advantage of his positions to increase the JIF of his own journal. Among the 82 manuscripts he handled as an editor and reviewer for other journals, he suggested 622 additional references, almost exclusively to the journal of which he was Editor-in-Chief (Davis, 2017a). The result was an astronomic rise in the JIF of the journal he edited, from 3.089 to 8.145 between 2014 and 2015. These flagrant abuses signal that editors are highly aware of the benefits derived from these manipulations.

Coercive self-citation is easier to identify than citation-stacking, which has become synonymous with the contemporary notion of "citation cartels". There can be several legitimate explanations for tightly coupled exchange of citations between journals, particularly in highly specialized fields. However, when these exchanges are done with the explicit intent of increasing the citedness of the journal, these are referred to as citation cartels. Although there have been a few attempts to identify cartels (Davis, 2012; Mongeon, Waltman, de Rijcke, 2016; Heneberg, 2016), detection is difficult on a number of fronts. Technically, the ability to identify cartels becomes more difficult as the size of the cartel increases. Furthermore, the notion of a cartel implies intentionality and premeditation—something that is impossible to prove using bibliometric data alone.

Thomson Reuters (and, subsequently, Clarivate Analytics) has worked to police inappropriate citation activity—though they note that they do not "assume motive on behalf of any party" (Clarivate Analytics, 2017). Each year, Clarivate provides a report of titles suppressed due to "anomalous citations patterns" and the reason for removal from the JCR (e.g., Thomson Reuters, 2016). Journals can be removed due to excessive self-citation or citation stacking, although thresholds are considered to be "extremely high" (Davis, 2017b). For example, in an analysis in 2002, the Institute for Scientific Information (the precursor to Thomson Reuters and Clarivate) found that for 82% of their titles, self-citation rates were at or below 20% (McVeigh, 2002). It is assumed, therefore, that all journals will engage to some degree in self-citation. However, when the proportional increase in the JIF is due largely to an increase in self-citation, the journal is flagged for further analysis (Clarivate Analytics, 2017). This is not an entirely uncommon practice and represents the dominant reason for suppression from the JCR. Perhaps as a result of reporting, cases of citation stacking have decreased over time (Hubbard, 2016).

Other scholars have also sought to create indicators for identifying excessive self-citations: Chorus and Waltman (2016) created the Impact Factor Biased Self-citation Practices (IFBSCP) indicator to examine the relationship between the share of self-citations for the years included in the impact factor to those in the preceding five years. To validate this as an indicator of coercive self-citations, they examined the rates of IFBSCP for the 64 journals identified in Wilhite and Fong (2012) as engaging in coercive citation behavior. They found that the named journals had IFBSCP rates 25% higher than the average Social Science journal, which suggests that their indicator measure is related with coercive behavior. This suggests that indicators may be developed to help identify—and hopefully curb—inappropriate citation behavior.

## 4.2 Role of evaluation policies

Impact Factor engineering does not happen in a void: these actions are a consequence of evaluation policies and practices. Institutions and individuals are complicit actors in promoting the JIF in a research evaluation context. Although soft persuasions towards maximizing impact can be seen across the scientific system, they are made most manifest in the cash-based reward systems, such as those documented and publicized in China. Chinese policies offering financial reward based on WOS-indexed publications began in earnest in the 1990s, to motivate production and increase international visibility (Quan, Chen, Shu, 2017). However, as noted by other studies (Butler, 2003), increasing national production does not necessarily equate to an increase in citedness, and might actually lead to a decrease. Therefore, China has moved steadily away from publication-based incentives in favor of citation-based indicators, particularly those based on JCR-quartiles of JIFs (Quan, Chen, Shu, 2017). At face value, these policies seem well-intentioned and even laudable—encouraging quality of quantity. However, given that the cash award for a *Nature* or *Science* article can be 20 times an annual salary in China (Quan, Chen, Shu, 2017), these rewards can create strong incentives for inappropriate behavior. Although one cannot determine causality, the rise in fraudulent authorship, data falsification and data fabrication in China (Qiu, 2010) in parallel with these rewards is disconcerting. There is even evidence of an industry of authorship for sale in China, in which authorship is sold to scholars at rates than often exceed salaries (Hvistendahl, 2013).

Furthermore, cash incentive programs have been correlated with increased submission, but not with publication (Franzoni, Scellato, & Stephan, 2011). Although most authors are fairly efficient at selecting appropriate journals, many authors tend to submit to higher impact factor journals first and then resubmit down the JIF ladder until they find an acceptance (Calcagno et al., 2012; Monastersky, 2005). Increasing the pressure to submit to high impact factor journals creates a burden on the scientific system and slows the pace of science as editors and reviewers are tasked with reviewing papers that are not submitted to the most appropriate venues. On a more fundamental level, financial rewards for papers externalizes the incentive to do scientific work. This contradicts central ideals of scholarship, in which scholars should be free from external pressures (Merton, 1973). A reward more than 20 times an annual salary inverts the reward system—prioritizing external (i.e., economic capital) over intrinsic (academic capital) rewards.

There is also a danger in tying rewards to publication in particular journals. The most appropriate venue for many scholars—particularly those in the social sciences and humanities—may not be in a WOS-indexed publication at all. By emphasizing JIFs, the coverage biases of the WOS become prioritized (Jin & Rousseau, 2004); that is, journal articles in the natural and medical sciences published in English are particularly incentivized. Some have argued that switching to English-language journals increases the visibility of science produced in countries where English is not the dominant language (e.g., Garfield, 1967; Cryanoski, 2010). However, others have expressed concern about the effects of a monolingual scholarly publishing industry (e.g., Shao & Shen, 2011). For instance, Larivière (2014) has shown that Canadian scholars in the social sciences and humanities were three times less likely to publish on Canada-related research topics when publishing in US journals than in Canadian journals, which demonstrates how journal venues directly affect the type of research performed.

**4.3 Application at the individual level**

As the JIF is based on a skewed distribution and, thus, is a weak predictor of individual papers' citation rates, its use as an indicator of the "quality" of individual researchers and papers—sometimes labelled as the ecological fallacy (Alberts, 2013)—is perhaps the most egregious misappropriation of the indicator. As Anthony van Raan noted: "if there is one thing every bibliometrician agrees, it is that you should never use the JIF to evaluate research performance for an article or for an individual—that is a mortal sin" (quoted in Van Noorden, 2010, p. 864-865). A less hyperbolic, but similarly unequivocal statement can be found from other bibliometricians: Henk Moed noted that such measures "have no value in assessing individual scientists" (2002). Despite these admonitions, the JIF is increasingly used as an indicator to evaluate individual scholars (see, among others, Quan, Chen & Shu, 2017; Rushforth & de Rijcke, 2015). While some might argue that publication in a high JIF is itself an achievement, given the relatively lower acceptance rates of these journals, the concern is more about the equation of the value of an article or individual with the past ranking of a journal (Brumback, 2012). This can lead to gross goal displacement (Osterloh & Frey, 2014), in which scholars tailor their topics for certain indicators.

Scholars are increasingly "thinking with indicators"—that is, allowing indicators to guide the process of science-making (Müller & De Rijcke, 2017). Specifically, scholars choose topics and dissemination venues not on scientific bases, but rather to meet certain incentive structures. In doing so, scholars substitute a "taste for science" with a "taste for rankings" (Osterloh & Frey, 2014). This is not a particularly novel claim. As early as 1991, Holub and colleagues noted that "WHERE a scientist published has become much more important than WHAT he is publishing" (*capitalization in original*). However, the impact factor obsession (Hicks, et al., 2015) has grown to the level where some scholars would rather destroy a paper than publish below a certain JIF threshold (Shibayama & Baba, 2015). This has led to a complicated and cyclical relationship between JIF, value, and reputation that is increasingly internalized into the process of scholarship (Müller & De Rijcke, 2017).

Scholars are aware of these negative effects: several initiatives in recent years have sought to disentangle journal rankings from individual rankings. At the 2012 annual meeting of the American Society for Cell Biology, a group of editors and publishers produced the San Francisco Declaration on Research Assessment, colloquially referred to as DORA (ASCB, 2012). The declaration called for the elimination of the use of JIFs for assessment of individual scholars and articles (ASCB, 2012), stating that the JIF was not appropriate "as a surrogate measure of the quality of individual research articles, to assess an individual scientist's contribution, or in hiring, promotion or funding decisions" (2012, 2). As of July of 2017, the declaration had nearly 13,000 individual signers and nearly 900 organizational signers. Funding agencies have also responded: the National Health and Medical Research Council in Australia produced a statement unequivocally denouncing JIFs for evaluating individual papers (NHMRC, 2010) and discontinued reporting of JIFs for evaluation. Nobel laureates and other high profile scholars have also spoken out against JIFs (Doherty, 2015) and boycotted high impact factor journals (Sample, 2013). However, these are privileged boycotts and resistance is much more difficult for those who are not well-established in the scientific system.

**4.4 Knock-off indicators**

The JIF has become a brand and, like any other luxury good, there is an industry of imitation. In recent years, a cottage industry of fake impact factors has emerged, with strong ties to predatory publishers. Librarian Geoffrey Beall—who for many years ran the well-known and controversial list of predatory publishers—identified more than 50 organizations that provide "questionable" or "misleading" metrics at the researcher, article, and journal level (Beall, 2017). The complicated web of mimicry is difficult to disentangle: the names of the organizations often replicate the name or acronym of the Institute for Scientific Information—e.g., the *Institute for Science Information (ISI)*, the *Index Scientific Journals (ISJ)*, or the *International Scientific Indexing (ISI)*—or the JIF—e.g., the *Journal Influence Factor-JIF*, the *General Impact Factor*, or the *Science Impact Factor*. One organization even goes as far as to imitate both the name of the indicator and that of the organization: journals can apply to the "Global Institute for Scientific Information" (GISI) to obtain a "Journal Impact Factor" (Global Institute for Scientific Information, 2017). Several journals seem to have either fallen prey or are complicit in this deceit: for instance, the list of journals to which GISI has attributed a "Journal Impact Factor" increased from 24 in 2010 to a high of 668 in 2011-2013. The numbers have been steadily dwindling, but there are still 153 journals listed in 2016. The listed journals come from both predatory and well-established publishers.

The organizations often go to lengths to maintain their deceit. For example, one website includes a red pop-up box warning editors and publishers that another company is scamming the original predatory company. The text reads: "This is to inform you that somebody is using our name (International Impact Factor Services) to deposit the fee for Impact Factor &he saying that he show your impact factor in our website, but do not reply those mails. If you answer those mails you will responsible for that" (International Impact Factor Services, 2017). This is not the only bait and switch in the impact factor market. For example, one of the only published articles on fake JIFs was published in *Electronic Physician: Excellence in Constructive Peer Review* (Jalalian, 2015). This article provides an account of so-called "bogus" indicators such as the Universal Impact Factor (UIF), Global Impact Factor (GIF), and Citefactor. The article describes the threat of these indicators to reputable indicators such as Thomson Reuters *and* the Index Copernicus metric value (ICV). However, the ICV, which is prominently displayed on the website of the *Electronic Physician*, is itself under scrutiny for its association with predatory journals (Beall, 2013). Therefore, this article seems to provide much the same function as the pop-up box of the International Impact Factor Services: It is a classic redirect technique, wherein the service attempts to legitimatize their own activities by delegitimizing others.

One of the biggest concerns with these products is the lack of transparency in the compilation of the indicators. The Global Impact Factor obliquely combines some for of peer review with the number of papers published (Global Impact Factor, 2017). Journals of the "Academy of IRMBR International Research in Management and Business Realities"—contained in Beall's list—rely on GoogleScholar to generate indicators (Academy of IRMBR, 2017), which seems a common approach for these fake JIFs. While one could argue that many of these indicators are legitimate competitors, rather than exploitative knock-offs, the mimicry of the names and acronyms as well as the cost structure begs caution. For example, the Global Impact Factor provides their indicator for an annual fee of $40 (Global Impact Factor, 2017) and *International Scientific Indexing* charges $100-130 per journal for the indicator and indexation on their platform (International Scientific Indexing, 2017). While the deceptive character of these sites might be apparent to many scholars, some have chosen to take a more neutral stance. For instance, a US university library guide on journal indicators lists these indicators alongside the JIF and other established indicators (Cal State East Bay, 2017). Other libraries have taken a more direct stance, urging their audience caution with these indicators and predatory publishers (George Washington University, 2017).

**5. What are the alternatives?**
Knock-off indicators abound, but there are also several other indicators that have emerged as complementary to or competitive with the JIF. This section examines four of the most established: the group of Eigenfactor Metrics, Source Normalized Impact per Paper (SNIP), CiteScore, and SCImago Journal Rank (SJR).

The Eigenfactor Metrics were introduced in 2010 as a new approach for ranking journals (West, Bergstrom, & Bergstrom, 2010). The metrics include two related indicators—the Eigenfactor Score and Article Influence Score—both based on the Eigenfactor algorithm, which leverages the citation network to identify and weight citations from central journals. The underlying algorithm is derived from Phillip Bonacich's (1972) eigenvector centrality, which has been employed across several domains, most notably as the foundation for Google's PageRank algorithm. The Eigenfactor Score depicts the "total value" of a journal and is thus size-dependent—as the size of the journal increases, so too will the Eigenfactor Score. The Article Influence Score, however, measures the average influence of articles in the journal, and is therefore more comparable to the JIF. However, there are several important differences: the AIS is calculated over a five-year (rather than two-year) time window, excludes self-citations, and uses weighted citations. Like the JIF, both indicators rely on Web of Science (WoS) data and were added to the JCR in 2009. As such, they represent a supplement to the JCR portfolio, rather than direct competition.

Scopus—the largest competitor to Web of Science—also has several associated journal indicators. The Source Normalized Impact per Paper (SNIP) indicator was proposed in 2009 by Henk F. Moed, then at the Centre for Science and Technology Studies (CWTS) of Leiden University (Moed, 2010) and later revised by Waltman and colleagues (Waltman et al., 2013). As discussed, one of the central interpretive critiques of the JIF is the inability to make cross-disciplinary comparisons. SNIP was developed to account for the different "citation potential" among fields. Rather than using an *a priori* journal-based classification, fields are defined according to the set of citing papers. In this way, the indicator is based on "contextual", rather than absolute, citation impact. Furthermore, SNIP serves to address another limitation of the JIF: by focusing on the set of citing papers, there is no concern about the asymmetries created by non-citable items. However, like the JIF, self-citations are included, which can lead to distortions in extreme cases. Furthermore, SNIP tends to be higher in journals with a large proportions of review articles, which causes additional bias. SNIP uses a three-year citation window—one year more than the JIF, but two less than the Article Influence Score.

Another indicator contained in "the Scopus basket of journal metrics" (Zijlstra & McCullough, 2016) is the SCImago Journal Rank (SJR), which was developed and continues to be updated by the SCImago research group at the University of Granada (González-Pereira, Guerrero-Bote, & Moya-Anegón, 2010). Like the Eigenfactor Score, the SJR employs Bonacich's eigenvector centrality to calculate the prestige of a journal, weighting the links according to the closeness of co-citation relationships (on the basis of citable documents). The current version of the indicator uses a three-year window, in keeping with the other Scopus journal indicators (Guerrero-Bote & Moya-Anegón, 2012). Furthermore, several heuristics are applied to circumvent gaming and distortions: in generating the prestige of the journal, there are thresholds on how much a single journal and the journal itself can provide—protecting against citation cartels and self-citations—and prestige is calculated on the basis of proportions rather than number of citable documents, to control for size and the dynamicity of the database.

In 2016, Elsevier released a new journal impact indicator, by the name CiteScore (Zijlstra & McCullough, 2016). The indicator is obtained by averaging, for a given journal, the number of citations received in a single year by papers it published during the preceding three years. The appeal is the simplicity—it is merely an average of citations received for all document types, which removes concerns about asymmetries between cited and citing items. However, the inclusion of all document types shifts the bias in another direction. While journals with a high proportion of non-citable items (e.g., editorials, news items) tend to fare well in the JIF, they are ranked lower in CiteScore. Critics of CiteScore have noted that this favorably biases Elsevier's own journals, which tend to publish a lower proportion of front matter than other journals (such as Nature's journals) (Bergstrom & West, 2016). Broader concerns have also been raised about the conflict of interest inherent in vertically integrated companies: There is considerable concern about the construction of indicators within a company that also publishes, indexes, and provides analytic services for journals (Sugimoto & Larivière, 2018). The increasing monopoly of Elsevier in this space has caused some to question the neutrality of the indicator.

However, none of these indicators have managed to displace the JIF's role in scientific system. The Eigenfactor Metrics are included in the JCR, but have not gained the marketing appeal of the JIF and the Scopus indicators have also not gained widespread traction after nearly a decade of existence. Part of this is the appeal of standardization: scholars working in research evaluation (whether hiring, promoting, or granting) have internalized the value of the JIF. Despite the well-known technical and interpretive concerns, the JIF remains the standard journal indicator.

**6. The future of journal impact indicators**
Building upon both original data and a review of the literature, this chapter provides a background for the creation of the JIF, an overview of its limitations, and a discussion of some of the most documented adverse effects. Several of the technical critiques can be or already are addressed by Clarivate. For instance, asymmetries between the numerator and denominator could be controlled by more careful analysis and cleaning of the data. Journal self-citation account for a minority of citations and can (and are already) flagged when excessive. The two-year JIF could be removed, in favor of a JIF with a longer citation window—which is already provided in more recent editions of the JCR. However, rather than replacing the original JIF with new indicators, these alternatives have merely been added to the JCR. This multiplicity of indicators is problematic from the perspective of standardization. When every researcher, administrator, evaluator, and policymaker is constructing tailor-made indicators, the indicators lose their central function—to communicate globally and across disciplines in a standard fashion (Sugimoto & Lariviere, 2018). Of course, bibliometrics is not alone in dedication to an imperfect indicator. For example, despite heavy criticism and the creation of alternative indicators (e.g., Bagust & Walley, 2000; Lebiedowska et al., 2008), the Body Mass Index remains, as per the World Health Organisation, the standard for the measurement of obesity.

However, some of the most disconcerting aspects are not purely technical, but rather due to the misapplication of the indicator. For example, one common technical concern is the skewness of citation distributions. Given that less than a third of articles are likely to achieve the citation value of the JIF, the indicator is misleading for application at the individual level. Due to skewness of citation distributions and the declining predictive power of the JIF, it is widely acknowledged that the indicator should not be used to evaluate individual articles or scholars (though there remains debate on this issue (Waltman & Traag, 2017)). Furthermore, the lack of normalization by discipline and the continual inflation of the indicator over time means that the JIF can only be used to rank contemporary journals within the same discipline.

It is also clear that it is not the indicator, but rather the application of the indicator that is causing systemic disruptions in science. Several of the adverse effects observed are not directly linked to JIF; rather, they are linked to the research evaluation system and, more specifically, to journals as vectors of scientific capital. In other words, the JIF has become synonymous with academic capital, and despite well-publicized criticisms (e.g., ASCB, 2012), it remains central to research evaluation. It would, of course, be naïve to assume that, in a pre-JIF era, there was no relationship between economic and scientific capital. Journals have long served at the heart of the race for scientific discovery: the certification and dissemination of knowledge allowed scholars to make priority claims, the traditional building blocks of scientific reputation (Dasgupta & David, 1994). However, the direct relationship between cash rewards and JIF is a gross perversion of the reward system in which economic incentives become the main objective of publishing. It is clear that measure has become the target (Strathern, 1997), as evident by the explicit manipulations within the system and the gross goal displacement in favor of high impact journals, whereby there is a prioritization of metrics over ethics (Franck, 1999; Osterloh & Frey, 2014).

When he published the first iteration of the JCR, Garfield (1976c) hoped that it could "prove itself indispensable to people who cannot rely on economic criteria alone in making basic decisions about journals, since the law of supply and demand is not always allowed to prevail" (p.1). The JIF became more than that: in many ways, it has become itself an economic item, capitalizing upon academic capital and the need for its measurement. As such, it has been grossly misapplied to make decisions about papers and authors, rather than journals, and caused distortions within the scholarly system. And while Garfield foresaw the use of the JIF for research evaluation, he also formulated recommendations for its proper use in his introduction of the first JCR (Garfield, 1976b, p.1):

> "Like any other tool, the JCR cannot be used indiscriminately. It is a source of highly valuable information, but that information must be used within a total framework proper to the decision to be made, the hypothesis to be examined, and rarely in isolation without consideration of other factors, objective and subjective."

Among these subjective factors, Garfield noted the reputation of the author, the controversial nature of the subject, the circulation and cost of the journal, and the degree to which the work is accessible. Garfield cautioned against comparing citation rates for journals in different disciplines and noted the biases in accounting for journals which do not use the Roman alphabet. While those factors remain quite relevant today, it seems they have been forgotten along the way. Moreover, since Garfield made these recommendations, English has become the *lingua franca* of research (Montgomery, 2013), which has led to a decline of the relative importance of non-English journals in many disciplines and, thus, reinforced the Web of Science—and, by extension, the JCR—as a measurement tool.

Despite these well-documented limitations and consequences, the JIF will likely remain part of the research ecosystem and as long as journals remain the primary mechanism for diffusing new knowledge, their reputation—as established by JIF or an alternative—will remain a marker of capital. It is essential, therefore, that actors within this system are provided with the means to interpret and apply the indicators responsibly, in full awareness of the consequences (Hicks et al., 2015; Lawrence, 2003). Perhaps more importantly, the scientific community must collectively ask: is the use of the Journal Impact Factor good for science?

**Cited References**